\documentclass[%
reprint,
superscriptaddress,
showkeys,
 amsmath,amssymb,
 aps,
prc,
]{revtex4-2}

\usepackage{graphicx}
\usepackage{dcolumn}
\usepackage{bm}
\usepackage{gensymb}
\usepackage{amssymb}
\usepackage{array}
\usepackage{textcmds} 
\usepackage{textcomp}
\usepackage{siunitx}

\newcommand{\msol}[1]{$M_{\odot}$}
\newcommand{\sfac}[1]{$S$-factor}

\begin{document}


\title{Refining the deep sub-barrier $^{12}$C + $^{12}$C excitation function with STELLA}


\author{J. Nippert}
\email[Corresponding Author:~]{jean.nippert@cea.fr}
\affiliation{Université de Strasbourg, CNRS, IPHC UMR 7178, 67000 Strasbourg, France}

\author{S. Courtin}
\email[Contact:~]{Sandrine.Courtin@iphc.cnrs.fr}
\affiliation{Université de Strasbourg, CNRS, IPHC UMR 7178, 67000 Strasbourg, France}
\affiliation{University of Strasbourg Institute of Advanced Studies (USIAS), Strasbourg, France}

\author{M. Heine}
\affiliation{Université de Strasbourg, CNRS, IPHC UMR 7178, 67000 Strasbourg, France}

\author{D.G. Jenkins}
\affiliation{University of York, York, YO10 5DD, UK}

\author{P. Adsley}
\affiliation{Institut de Physique Nucléaire, CNRS/IN2P3, Universit\'{e} Paris-Sud, Universit\'{e} Paris-Saclay, 91406 Orsay Cedex, France}

\author{A. Bonhomme}
\affiliation{Université de Strasbourg, CNRS, IPHC UMR 7178, 67000 Strasbourg, France}

\author{R. Canavan}
\affiliation{School of Mathematics and Physics, University of Surrey, Guildford, GU2 7XH, UK}
\affiliation{National Physical Laboratory, Teddington, Middlesex, TW110 LW, UK}

\author{D. Curien}
\affiliation{Université de Strasbourg, CNRS, IPHC UMR 7178, 67000 Strasbourg, France}

\author{T. Dumont}
\affiliation{Université de Strasbourg, CNRS, IPHC UMR 7178, 67000 Strasbourg, France}

\author{E. Gregor}
\affiliation{Université de Strasbourg, CNRS, IPHC UMR 7178, 67000 Strasbourg, France}

\author{G. Harmant}
\affiliation{Université de Strasbourg, CNRS, IPHC UMR 7178, 67000 Strasbourg, France}

\author{E. Monpribat}
\affiliation{Université de Strasbourg, CNRS, IPHC UMR 7178, 67000 Strasbourg, France}

\author{L. Morrison}
\affiliation{School of Mathematics and Physics, University of Surrey, Guildford, GU2 7XH, UK}

\author{M. Moukaddam}
\affiliation{Université de Strasbourg, CNRS, IPHC UMR 7178, 67000 Strasbourg, France}

\author{M. Richer}
\affiliation{Université de Strasbourg, CNRS, IPHC UMR 7178, 67000 Strasbourg, France}

\author{M. Rudigier}
\affiliation{School of Mathematics and Physics, University of Surrey, Guildford, GU2 7XH, UK}

\author{J.G. Vega Romero}
\affiliation{University of York, York, YO10 5DD, UK}

\author{W.N. Catford}
\affiliation{School of Mathematics and Physics, University of Surrey, Guildford, GU2 7XH, UK}

\author{P. Cotte}
\affiliation{Univeristé Paris-Saclay, IJCLab, CNRS/IN2P3, F-91405 Orsay, France}

\author{S. Della Negra}
\affiliation{Univeristé Paris-Saclay, IJCLab, CNRS/IN2P3, F-91405 Orsay, France}

\author{G. Haefner}
\affiliation{Univeristé Paris-Saclay, IJCLab, CNRS/IN2P3, F-91405 Orsay, France}

\author{F. Hammache}
\affiliation{Univeristé Paris-Saclay, IJCLab, CNRS/IN2P3, F-91405 Orsay, France}

\author{J. Lesrel}
\affiliation{Univeristé Paris-Saclay, IJCLab, CNRS/IN2P3, F-91405 Orsay, France}

\author{S. Pascu}
\affiliation{School of Mathematics and Physics, University of Surrey, Guildford, GU2 7XH, UK}
\affiliation{National Institute for Physics and Nuclear Engineering, R-77125, Bucharest-Magurele, Romania}

\author{Zs. Podoly\'{a}k}
\affiliation{School of Mathematics and Physics, University of Surrey, Guildford, GU2 7XH, UK}

\author{P.H. Regan}
\affiliation{School of Mathematics and Physics, University of Surrey, Guildford, GU2 7XH, UK}
\affiliation{National Physical Laboratory, Teddington, Middlesex, TW110 LW, UK}

\author{I. Ribaud}
\affiliation{Univeristé Paris-Saclay, IJCLab, CNRS/IN2P3, F-91405 Orsay, France}

\author{N. de S\'{e}r\'{e}ville}
\affiliation{Univeristé Paris-Saclay, IJCLab, CNRS/IN2P3, F-91405 Orsay, France}

\author{C. Stodel}
\affiliation{GANIL, CEA/DSM-CNRS/IN2P3, Caen, F-14076, France}

\author{J. Vesi\'{c}}
\affiliation{GSI Helmholtzzentrum f\"ur Schwerionenforschung, D-64291 Darmstadt, Germany}
\affiliation{Jožef Stefan Institute, Jamova cesta 39, SI-1000 Ljubljana, Slovenia}

\collaboration{STELLA collaboration}
\noaffiliation

\date{\today}

\begin{abstract}
  \begin{description}
  \item[Background]
    The fusion excitation function of $^{12}$C+$^{12}$C contains many resonances from above the Coulomb barrier towards sub-barrier energies. These oscillations might be linked to molecular states in the compound nucleus $^{24}$Mg with drastic consequences for stellar carbon burning. Moreover, these structures render the common extrapolations from measurements into the astrophysics region of interest at deep sub-barrier energy extremely uncertain. 
  \item[Purpose]
    We have investigated the cross section around the lowest direct coincident gamma-particle measurements, where previously only limits could be established with the aim of obtaining a detailed description of the excitation function. We have furthermore analysed the ratio of extreme decay branching into the first excited state of daughter nuclei with alpha or proton emission at relative kinetic energies where previous measurements are in disagreement with each other.
  \item[Methods]
    The experiments were carried out at the Andromède accelerator, IJCLab Orsay (France), with carbon beam intensities of up to 3~p$\mu{}$A on thin rotating carbon target foils. Light charged particles and gamma decays were detected using silicon strip detectors and LaBr$_{3}$(Ce) crystals, respectively, in coincidence measurements with nanoseconds precision.
  \item[Results]
    We establish data points with highly improved accuracy at the high-energy tail of the lowest resonance detected so far in direct measurements, right in the astrophysics region of interest (RoI). The findings are in agreement with an earlier interpretation of the fusion excitation function composed of resonances on top of a global trend of empirically determined fusion hindrance behaviour. At a relative energy between 3.1~MeV and 3.3~MeV, an oscillating behaviour of the branching ratio into the first excited final state with alpha or proton emission is observed, with extreme values significantly deviating from data at higher energy.
  \item[Conclusions] 
    Our findings in the astrophysics RoI support reaction-rate models with a lower average \sfac{} trend, that deviates significantly from standard extrapolations between 2.2~MeV and 2.6~MeV, for stellar carbon burning simulations of up to 25~\msol{} stars. Based on our data, an overall increase of the \sfac{} at deep sub-barrier energy cannot be confirmed. The extremely low ratio of the branching into the first excited state with proton over alpha emission of $\approx{}2\%$ at 3.23~MeV might indicate the presence of alpha cluster compound states in $^{24}$Mg. This highly favours $\alpha{}$ emission with fundamental consequences in possible stellar carbon burning sites.
  \end{description}
\end{abstract}


\maketitle


\section{\label{sec:Intro}Introduction}

The $^{12}$C+$^{12}$C fusion excitation function exhibits strongly resonant behavior from the region above the Coulomb barrier down to the deep sub-barrier energy regime relevant to nuclear astrophysics~\cite{almqvist1960, aguilera2006PRC, Spillane2007}. Those oscillations can provide information about nuclear structure and might be related to molecular states in the compound nucleus ~\cite{Ikeda1968, taras1978, chiba2015PRC}, that then strongly resemble a configuration of two carbon nuclei, with large overlap of the shape of the initial and final state wave function. It was furthermore pointed out, that in particular during deep sub-barrier fusion of identical spin-0 bosons, only a limited number of states, namely with positive parity, are accessible to the compound system~\cite{JiangPRL2013, jiang2021EPJA}. As a consequence, the cross section oscillation would reflect the sparse number of non-overlapping states of relevant angular momentum and appears to be an artifact of the systems degeneracy at low relative energy of the carbon nuclei.

Carbon fusion is a key step in the stellar nucleosynthesis of the elements and occurs in different phases of stellar evolution, explosive carbon burning in type Ia supernovae~\cite{woosley2002} and possibly in superbursts of x-ray binary systems~\cite{cumming2001}. Quiescent carbon burning takes place in cores of massive stars at temperatures of the order of 1~GK~\cite{rolfs1988cauldrons}, which translates into a Gamow window of a few MeV of the carbon nuclei and necessitates extrapolations of the direct measurements of the fusion cross section. In the astrophysics RoI at about $E_{\mathrm{rel}}\leq{}2.5$~MeV, the cross section drops below nano barn requiring sophisticated coincidence techniques for efficient background reduction~\cite{jiang2012, Fruet2020, tan2020PRL, tan2024PRC}. The reaction was also assessed by a \qq{Trojan Horse} experiment~\cite{tumino2018nat} by measuring the $^{12}$C($^{14}$Ne, $\alpha{}^{20}$Ne)d and  $^{12}$C($^{14}$Ne, p${}^{23}$Na)d three-body processes extracting the carbon fusion cross section assuming a deuterium spectator. These latter results predict an overall increase of the \sfac{} with decreasing energy. However, the trend obtained was suggested to be an artefact of an invalid plane-wave approximation during the interpretation of experimental data~\cite{mukhamedzhanov2019}.
The commonly used CF88 model from Fowler \textit{et al.}~\cite{Fowler1975} employs optical potential calculations and yields an \sfac{} that is systematically higher than experimental data in the deep sub-barrier regime. The empirical model by Jiang \textit{et al.}~\cite{JiangPRC2007} is based on a comprehensive study of fusing systems in this energy regime and predicts a broad \sfac{} maximum at $E_{\mathrm{rel}}=4$~MeV for $^{12}$C+$^{12}$C with a steep drop due to fusion hindrance towards deep sub-barrier energies~\cite{jiang2002PRL}, as compared to Ref.~\cite{Fowler1975} with deviations of orders of magnitude in the astrophysics RoI.
%
The impact of resonances and hindrance on stellar carbon burning was investigated in various studies pointing out the importance of precise knowledge or modeling of the reaction rate. In massive stars, low-lying resonances lead to lower central carbon burning temperature with significantly increased $s$-process production from more efficient $^{13}$C($\alpha$, n)~\cite{Pignatari2013}. Conversely, hindrance of carbon fusion leads to higher core temperature with shortened carbon burning lifetimes~\cite{GasquesPRC2007, Monpribat2022, dumont2024}. These works consider also the effects of deviations from the established branching ratio of the $\alpha$- and p-exit channels $R_{\alpha}/R_{p}$~\cite{bennett2012MNRAstrS}. An overall lowering/increase over the entire energy range affects $^{22}$Ne nucleosynthesis with lower/higher $s$-process production~\cite{Pignatari2013}. An extremely high $R_{\alpha}/R_{p}$ was detected at $E_{rel}=2.14$~MeV~\cite{Spillane2007}, which is still in reach of direct measurements. The effect on stellar evolution and nucleosynthesis in massive stars is, however, averaged out relative to the CF88 model~\cite{Monpribat2022}. An overall increase of the \sfac{} yields a significantly increased reaction rate~\cite{chieffi2021}, and affects the compactness of the star that might be linked to its explodibility, studied for a wide mass range of stars.
\section{\label{sec:exp}Experiment}
\begin{figure}[htp]
    \centering
    \includegraphics[width=0.5\columnwidth]{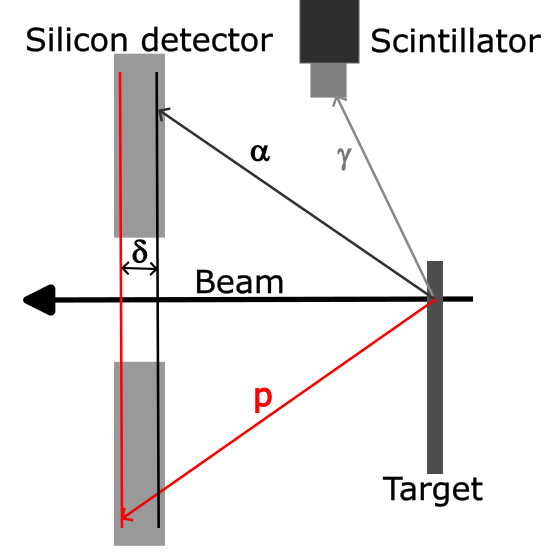}
    \caption{Coincidence setup with schematic of a silicon detector and a scintillator. The difference of the implantation depth $\delta{}\approx{}200$~$\mu{}$m at the particle energy of interest of alphas (black) and protons (red) is indicated.}
    \label{fig:timediffscheme}
\end{figure}

In this work, we present $^{12}$C+$^{12}$C cross-section measurements at energies from $E_{\mathrm{rel}}=2.32$ to 3.82~MeV using the STELLA (STELlar LAboratory)~\cite{HeineNIM2018} apparatus with UK-FATIMA (FAst TIMing Array)~\cite{roberts2014, rudigier2020}. In this energy regime, the comprehensive fusion excitation functions from Becker~\textit{et al.}~\cite{Becker1981} and Spillane~\textit{et al.}~\cite{Spillane2007} are diverging. The experimental campaign in Ref.~\cite{Becker1981} made use of thin targets detected light charged particles with comprehensive analysis of the angular emission distribution. The beam energy was lowered in steps of 50~keV above $E_{\mathrm{rel}}=3.5$~MeV and 100~keV below resulting in an overlap between subsequent measurements. In Ref.~\cite{Spillane2007}, de-excitation gammas of daughter nuclei in the final state were detected with branching ratio corrections from Ref.~\cite{Becker1981} in a thick target experiment with energy steps of 50~keV. The gamma energy associated to the exit channel with proton emission is much lower than for alpha emission inducing substantially higher uncertainty on the cross sections. In the energy region discussed, a resonant structure was measured for the alpha channel, but cannot be clearly resolved for the proton channel, but with huge discrepancies between both \sfac{} data sets.
The present experiment was performed at the 90$\degree$ beam line at the Andromède accelerator~\cite{ELLER2015367} (IJCLab Orsay, France), providing  $2^+/3^+$ carbon beam with intensity up to 3~p$\mu$A. Inside  the scattering chamber, self-supported large thin carbon targets with area density from 20 to 80~$\mu$g$/$cm$^{2}$, on rotating frames with an inner diameter of 46~mm were used for efficient heat dissipation with an adjusted rotation velocity of up to 1000~rpm. The pumping, the layout of the chamber and heating of the target from the elevated beam intensity prevents quite naturally against previously reported target thickening in such experiments~\cite{aguilera2006PRC, healy1997}. The vacuum inside the chamber is made with a dry pump combined to a cryogenic pump with a wide opening towards the chamber reaching a vacuum of $\approx 10^{-8}$~mbar. To complement the thickness determination right after production, a thickness measurement chamber has been developed. It employs the energy loss of an alpha emitting source that is measured on and off the beam spot after irradiation and compared to reference signals. No build up has been measured within the uncertainty of 10$\%$.
The carbon fusion reaction channels discussed here are 
\begin{align}
 ^{12}\mathrm{C}+^{12}\mathrm{C}\rightarrow{} ^{24}\mathrm{Mg}^{*} & \rightarrow{}^{20}\mathrm{Ne}^{*} + \alpha{}, \label{equ:fus_a} \\
 & \rightarrow{}^{23}\mathrm{Na}^{*} + \mathrm{p}, \label{equ:fus_p}
\end{align}
with emission of light charged particles and de-excitation gammas from excited final states, that are detected in coincidence for efficient background reduction being sensitive to cross-section measurements in the sub nano barn region~\cite{Fruet2020}. The final exit channel to the first excited state in $^{20}$Ne and $^{23}$Na is hereafter denoted with $\alpha_{1}$ and p$_{1}$, respectively.
The gamma detection array is composed of 36 $1.5"\times2"$ cylindrical LaBr$_3$(Ce) scintillators in a compact cylindrical configuration with a typical energy resolution of $3\%$ (FWHM) at 1333~keV and intrinsic timing resolution lower than nano second. The intrinsic radioactivity of $^{138}$La is used for automated energy calibration by matching experimental data with detailed simulations of the decay process. The width of the peaks at the gamma energies of interest ($E(\mathrm{p}_{1})=440$~keV, $E(\alpha{}_{1})=1634$~keV) was compared to calibrating with an $^{152}$Eu source and agrees within the energy resolution of the crystals. The full energy peak detection efficiency of this experiment is 8.6\% and 2.6\% at $E(\mathrm{p}_{1})$ and $E(\alpha{}_{1})$, respectively. The charged particles are detected inside of the reaction chamber using $500~\mu$m thick double sided striped silicon detectors covering 30$\%$ of 4$\pi$, one S3 type from Micron Semiconductor Ltd.\textregistered{} in the forward direction, one S1 and S3 type in the backward direction with respect to the beam. Figure~\ref{fig:timediffscheme},
displays the emitted charged particles and gammas during the coincidence measurements.
Elastic scattering of carbon beam is monitored at 45$\degree$ at a distance of 23~cm from the target, which, in combination with beam current measurements behind the target, yields reliable control of the state of the intact target foil.
\section{\label{sec:Ana}Analysis}
\begin{figure}[htp]
    \centering
    \includegraphics[width=0.8\columnwidth]{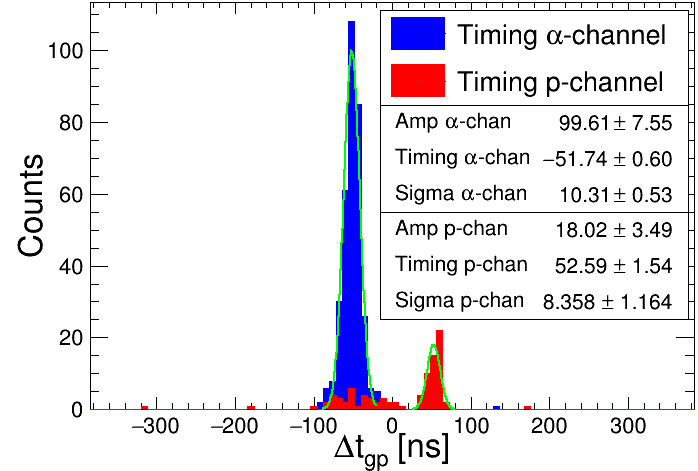}
    \caption{Gamma-particle time difference. The blue and red distribution represents alphas and protons with timing resolution of 10.3(5)~ns and 8(1)~ns, respectively, and the green curve is a Gaussian distribution fit.}
    \label{fig:timediff}
\end{figure}

\begin{figure}[htp]
    \centering
    \includegraphics[width=0.99\columnwidth]{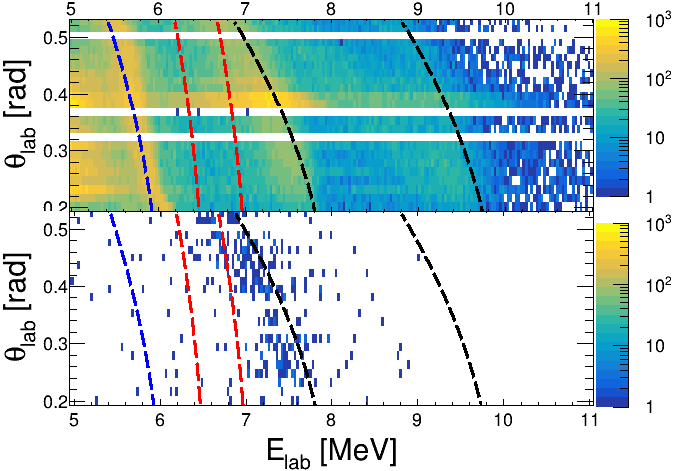}
    \caption{Polar angle \textit{vs.} particle energy at a relative energy $E_{\mathrm{rel}} = 3.28$~MeV. The red and black lines represent the nominal angular energy dependence of alphas and protons, respectively, and the blue line indicates the $^{12}$C$(d, p)^{13}$C background reaction. The upper panel shows raw particle data while the lower panel is for events where a 1634~keV gamma ray, as expected for the $\alpha_1$-channel, was detected in coincidence.}
    \label{fig:2Dene}
\end{figure}

During coincident data taking, the energies of light charged particles and gamma rays are recorded with associated time stamps. The combination of the 125~MHz (STELLA) and the 1~GHz (UK-FATIMA) data acquisition allows for nanoseconds event selection and the impact on extracting of the exit channels is illustrated in Figure~\ref{fig:timediff},
where the time difference between gammas and particles (alphas in blue, protons in red) is shown. Both distributions can be selected with high accuracy in the time domain and are well separated from each other. The indicated proton counts around $t_{\mathrm{gp}}\approx{}-50$~ns on top of the alpha timing distribution arise from selecting the according gamma and proton energy. These events can, however, be identified as alpha background, namely from Compton events of the higher energy $E(\alpha_1)$, that also fall in the full energy selection gate of $E(\mathrm{p}_1)$ in the LaBr$_3$(Ce) crystals, in the nanoseconds time domain displayed in Figure~\ref{fig:timediff}.
The coincidence timing resolution (sigma) for alphas and protons is about 10~ns and 8~ns, respectively. This was achieved by selecting events from final states with alphas, in an iterative loop over for the entire S3 detectors, and matching the time response of neighbouring strips for events with shared charge between strips. The same offsets are obtained independently by using gamma-particle events adjusting each strip separately. This calibration, on the other hand, is not necessary for final states with proton emission. The situation is illustrated in Figure~\ref{fig:timediffscheme}, where the implantation depth of alphas (black) and protons (red) in the detector substrate (gray) is indicated. The difference of the penetration depth $\delta{}$ is caused by the lower ionisation potential of protons in the silicon material. In addition, the localisation in depth of the Bragg peak is more diffuse for alphas~\cite{ziegler2010} (reflecting also in the rise time of the electronics signal) with lower gamma-particle timing resolution. While for 7~MeV and 10~MeV alphas, the penetration depth in silicon is 40~$\mu{}$m and 69~$\mu{}$m, respectively, for 6~MeV and 7~MeV protons, the penetration depth in silicon is 295~$\mu{}$m and 384~$\mu{}$m, respectively. As a consequence, protons reach further towards the readout electrode causing an offset due to the shorter drifting path of the induced charge electrons with respect to alphas, but with similar timing in all strips.
The event selection for carbon fusion cross sections in the current analysis comprises gating on the gamma and particle energy, and in addition on the associated timing interval. Figure~\ref{fig:2Dene}
shows how kinematic loci due to evaporated protons and alpha particles are identified, with the polar angle \textit{versus} the energy of the charged particles at a relative energy $E_{\mathrm{rel}} = 3.28$~MeV. The black and red lines represent the nominal particle energy of the alpha and the proton channel, respectively, where either way the higher energy is related to ground state transitions and the lower energy to final states in the first excited state. The blue line is associated to the dominating $^{12}$C$(d, p)^{13}$C background reaction, that arises from air humidity that condensed on the target foil and yields proton background in the RoI of fusion reactions~\cite{zickefoose2018}. 
The upper panel in Figure~\ref{fig:2Dene} shows the charged particle spectrum without coincident gamma selection, while for the lower panel a gamma energy $E(\alpha_1)=1634$~keV is required. The coincidence condition significantly cleans up the spectrum for reliable fusion event selection. Possible random coincidences were accounted for by subtracting spectra gated on the non-coincident timing regime (see Ref.~\cite{heine2022EPJWeb} for details).
At the lowest energy $E_{\mathrm{rel}} = 2.4$~MeV, fusion cross sections in the order of nanobarn can be expected. Therefore, the approach of statistical analysis by Feldman and Cousins~\cite{feldman1998} for proper definition of the confidence intervals with low counts was applied, likewise done in Ref.~\cite{Fruet2020}.
The counts were then normalised to the beam intensity and the target thickness taking into account the detection efficiency and angular coverage of the gamma and particle detectors. The measurements of the final exit channels $\alpha_1$ and p$_1$ were converted into total alpha and proton cross sections with branching corrections extracted from Becker~\textit{et al.} ($\alpha_1/\alpha=0.319$, p$_1$/p=0.156) averaged over all energies in Ref.~\cite{Becker1981}. The systematic uncertainty of 16.5\% comprises the beam intensity, target thickness, solid angle of the detectors, gamma detection efficiency and branching ratio correction. The cross sections are given with statistical uncertainties, and were converted into the modified \sfac{},
\begin{equation}
    S^* = \sigma E \exp(2\pi\eta + gE),
\end{equation}
with $\eta = Z_1Z_2e^2/(\hslash\nu)$ the Sommerfeld parameter where $Z_1$ and $Z_2$ are the charge of the nuclei and $e$ the electron charge and $g = 0.122\sqrt{\mu R^3/(Z_1Z_2)}$ the form factor for angular momentum $l = 0$ states in $^{12}$C+$^{12}$C fusion reactions using a square-well potential for a reduced mass $\mu$ \cite{Patterson1969,rolfs1988cauldrons}. The results are visualised in the form of the astrophysical \sfac{} instead of cross section, because the former has the gross sub-barrier energy dependence factored out and thus better shows nuclear physics aspects such as resonant behaviour or sub-barrier hindrance.
The beam undergoes energy loss in the target foil from multiple scattering from the target atoms. For the calculation of the effective relative energy $E_{\mathrm{rel}}$, the cross section decrease within the target thickness is approximated by a constant yielding an $E_{\mathrm{rel}}$ at the center of the target. The values for both $E_{\mathrm{beam}}$ and $E_{\mathrm{rel}}$ are summarized in Table~\ref{tab:XS} alongside the total alpha and proton cross sections.
\begin{table*}[htp]
    \caption{\label{tab:XS} $E_{\mathrm{b}}$ and $E_\mathrm{rel}$ are the nominal beam energy and the effective energy in the center-of-mass system, respectively (see text for details). The values in the laboratory system are twice as big. The cross-sections for the alpha and proton channel are given, as well as the ratio $\sigma_{p1}/(\sigma_{\alpha{}1} + \sigma_{\mathrm{p}1})$.}
        \begin{tabular}{ll r@{$\,\pm\,$}l r@{$\,\pm\,$}l r@{$\,\pm\,$}l}
            \hline \hline
            & & \multicolumn{2}{c}{} & \multicolumn{2}{c}{} & \multicolumn{2}{c}{} \\
            $E_{\mathrm{b}}$ [MeV] & $E_\mathrm{rel}$ [MeV] & \multicolumn{2}{c}{$\sigma_{\mathrm{p}}$ [mb]} & \multicolumn{2}{c}{$\sigma_\alpha$ [mb]} & \multicolumn{2}{c}{$\sigma_{p1}/(\sigma_{\alpha{}1} + \sigma_{\mathrm{p}1})$ [\%]} \\
            & & \multicolumn{2}{c}{} & \multicolumn{2}{c}{} & \multicolumn{2}{c}{} \\
           \hline
            & & \multicolumn{2}{c}{} & \multicolumn{2}{c}{} & \multicolumn{2}{c}{} \\
           3.47 & 3.37 & $(5.2$ & $0.2)\mathrm{x}{}10^{-4}$ & $\quad{}(1.90$ & $0.05)\mathrm{x}{}10^{-3}$ & $\quad{}11.8$ & 0.3 \\
           3.40 & 3.28 & $(1.5$ & $0.3)\mathrm{x}{}10^{-4}$ & $\quad{}(1.29$ & $0.07)\mathrm{x}{}10^{-3}$ & $\quad{}5.41 $ & 0.08 \\
           3.30 & 3.23 & $(3.0$ & $0.9)\mathrm{x}{}10^{-5}$ & $\quad{}(8.6$  & $0.5)\mathrm{x}{}10^{-4}$  & $\quad{}1.68 $ & 0.01 \\
           3.30 & 3.13 & $(6.6$ & $0.9)\mathrm{x}{}10^{-5}$ & $\quad{}(4.5$  & $0.2)\mathrm{x}{}10^{-4}$  & $\quad{}6.6 $ & 0.1 \\
           3.20 & 3.10 & $(8.4$ & $0.7)\mathrm{x}{}10^{-5}$ & $\quad{}(2.9$  & $0.2)\mathrm{x}{}10^{-4}$  & $\quad{}12.5$ & 0.4 \\
           3.13 & 3.05 & $(3.4$ & $0.6)\mathrm{x}{}10^{-5}$ & $\quad{}(2.3$  & $0.2)\mathrm{x}{}10^{-4}$  & $\quad{}6.5 $ & 0.1 \\
           2.41 & 2.32 & $(1.7$ & $1.4)\mathrm{x}{}10^{-7}$ & $\quad{}(2.5$  & $1.8)\mathrm{x}{}10^{-7}$  & $\quad{}24.5$ & 4.0 \\
           & & \multicolumn{2}{c}{} & \multicolumn{2}{c}{} & \multicolumn{2}{c}{} \\
           \hline \hline
        \end{tabular}
\end{table*}
\section{\label{sec:res}Results}
\begin{figure}[htp]
    \centering
    \includegraphics[width=0.99\columnwidth]{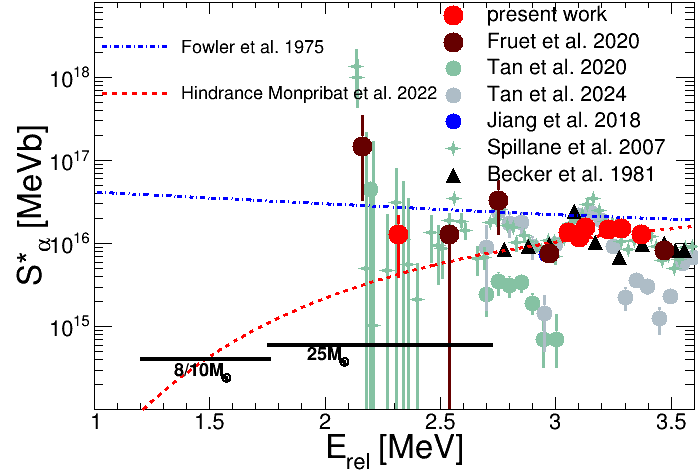}
    \caption{Modified \sfac{} for $^{12}$C+$^{12}$C as a function of the effective relative energy for final exit channels with alpha emission. The present data are compared to experiments with gamma detection (crosses)~\cite{Spillane2007}, particle detection (triangles)~\cite{Becker1981} or the coincidence technique (circles)~\cite{jiang2018PRC, Fruet2020, tan2020PRL, tan2024PRC}. The blue dash-dotted line represents the commonly used extrapolation of the data~\cite{Fowler1975}, while the red dashed line corresponds to a sub-barrier fusion suppression model~\cite{JiangPRC2007}. The marked Gamow windows correspond to 0.5 and 0.9~GK for 8-10~\msol{} and 25~\msol{}, respectively~\cite{Iliadis}.}
    \label{fig:sfaca}
\end{figure}

\begin{figure}[htp]
    \centering
    \includegraphics[width=0.99\columnwidth]{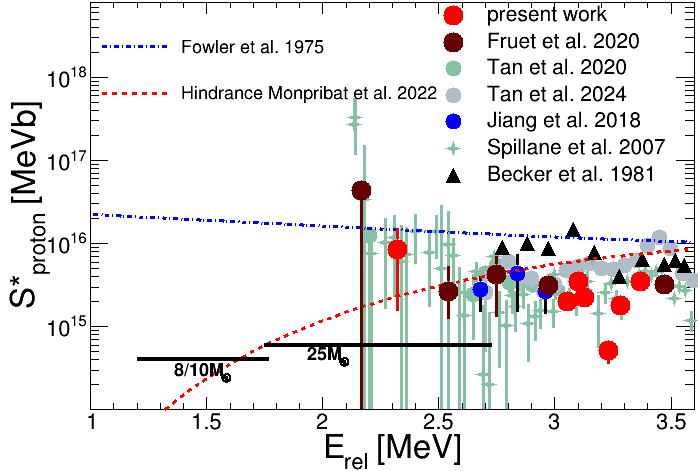}
    \caption{Modified \sfac{} for $^{12}$C+$^{12}$C as a function of the effective relative energy for final exit channels with proton emission. The present data are compared to experiments with gamma detection (crosses)~\cite{Spillane2007}, particle detection (triangles)~\cite{Becker1981} or the coincidence technique (circles)~\cite{jiang2018PRC, Fruet2020, tan2020PRL, tan2024PRC}. The blue dash-dotted line represents the commonly used extrapolation of the data~\cite{Fowler1975}, while the red dashed line corresponds to a sub-barrier fusion suppression model~\cite{JiangPRC2007}. The marked Gamow windows correspond to 0.5 and 0.9~GK for 8-10~\msol{} and 25~\msol{}, respectively~\cite{Iliadis}.}
    \label{fig:sfacp}
\end{figure}

\begin{figure}[htp]
    \centering
    \includegraphics[width=0.99\columnwidth]{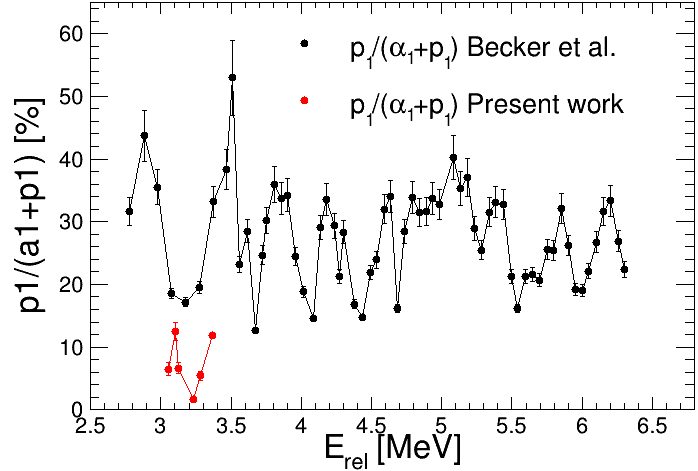}
    \caption{Cross section ratio $\sigma_{p1}/(\sigma_{\alpha{}1} + \sigma_{\mathrm{p}1})$ in percent of the present data (red) in comparison to Becker~\textit{et al.}~\cite{Becker1981} (black).}
    \label{fig:fracBeck}
\end{figure}

The \sfac{} for $^{12}$C+$^{12}$C fusion cross sections corresponding to the exit channel with alpha emission are presented in Figure~\ref{fig:sfaca},
and with proton emission in Figure~\ref{fig:sfacp}
as a function of relative energy. The present data are compared to comprehensive campaigns~\cite{Becker1981, Spillane2007} and coincidence experiments~\cite{jiang2018PRC, Fruet2020, tan2020PRL, tan2024PRC}. The extrapolations correspond to an average cross section (CF88, blue dash-dotted line) predicting a moderate \sfac{} increase towards lower energy~\cite{Fowler1975} and a phenomenological hindrance model (red dotted line) with a steep \sfac{} drop~\cite{JiangPRC2007}. The \sfac{} with alpha emission to the first excited state of $^{20}$Ne ($\alpha_{1})$ and with proton emission to the first excited state of $^{23}$Na (p$_{1})$ are presented in the supplementary material~\cite{nippert2024_suppl}. The current data were taken in two different energy regimes with several measurements between $E_{\mathrm{rel}}=3.05\dots{}3.37$~MeV and a data point at $E_{\mathrm{rel}}=2.32$~MeV between the energies of two earlier measurements with the STELLA setup. For some of those, only upper limits instead of data points could be established (see supplements in Ref.~\cite{Fruet2020}).
Our data present higher precision as compared to conventional measurements in particular in the low-energy region, where our results appear to agree with the hindrance model aligning with earlier coincidence technique experiments, while being slightly below the CF88. This finding supports the interpretation of the excitation function in this energy region as a superposition of an underlying hindrance trend with a narrow resonance at $E_{\mathrm{rel}}=2.14$~MeV, that was first reported by Spillane~\textit{et al.}, and used in the analysis of Fruet~\textit{et al.} (green crosses and brown circles, respectively, in Figure~\ref{fig:sfaca} and \ref{fig:sfacp}).
The measurements at the higher energy range are motivated by the diverging of the data sets from Becker~\textit{et al.}~\cite{Becker1981} and Spillane~\textit{et al.}~\cite{Spillane2007} below $E_{\mathrm{rel}}=3.5$~MeV, most prominently apparent in the proton exit channel, while in the alpha channel \sfac{} a resonant structure was reported, although this appears offset by around 100~keV compared to the data presented in Ref.~\cite{Becker1981}. On top of that, the statistical uncertainty is increasing caused by cross sections of $10^{-3}$~mb and below. In the current work, we performed coincidence measurements of the $\alpha_1$ and p$_1$ final states and scaled to the absolute \sfac{}. Most evidently, the uncertainty in the proton channel (Figure~\ref{fig:sfacp}) is much improved. Data are in the \sfac{} range of Ref.~\cite{Spillane2007}, but reveal a much steeper drop with a minimum at $E_{\mathrm{rel}}=3.23$~MeV, also compared to Ref.~\cite{Becker1981}, where only a slight dip is reported but at higher \sfac{} values. We do not observe a particular structure of the alpha channel in this energy region, but rather average over the peak reported in Ref.~\cite{Spillane2007}.
To address the sensitivity of the total cross section to the partial cross section, we present the fraction $\sigma_{p1}/(\sigma_{\alpha{}1} + \sigma_{\mathrm{p}1})$ in percent in Table~\ref{tab:XS} and Figure~\ref{fig:fracBeck}
in comparison to the study by Becker~\textit{et al.}. The present data (red) deviate significantly from this systematic (black) displaying extreme p$_1$ proton suppression of down to $1.7\,\%$ at $E_{\mathrm{rel}}=3.23$~MeV. We note that a dip of the proton \sfac{} is also observed in Ref.~\cite{Becker1981, Spillane2007} in this energy region. The comparison to Spillane~\textit{et al.} as carried out in Figure~\ref{fig:fracBeck} would depend on details about the detected lines and the applied branching correction.
\section{\label{sec:ccl}Discussion and Conclusion}
While the proton \sfac{} data at $E_{\mathrm{rel}}=2.32$~MeV lies between the CF88 and hindrance trend, the alpha \sfac{} rather supports modelling of the excitation function with fusion hindrance and a narrow resonance at $E_{\mathrm{rel}}=2.14$~MeV (HinRes). In this picture, the present data are situated at the high-energy tail of the resonance. Overall, the coincidence data tend not to carry a brisk increase of the \sfac{} in this energy region. This finding is in agreement with coincidence measurements in Ref.~\cite{tan2024PRC} and a broader analysis over a wider energy region. Data are rather well approximated by HinRes in the energy region around the lowest direct measurements. This model was utilized in Ref.~\cite{Monpribat2022} for carbon burning simulations in 12~\msol{} and 25~\msol{} stars that takes place at core temperatures corresponding to this energy regime. Further measurements with high accuracy are therefore desired to improve the input fusion excitation function for such stellar simulations. In particular, precise direct measurements at the low-energy tail of the resonance at $E_{\mathrm{rel}}=2.14$~MeV, where a cross section of a few pico barn is expected, would be striking for the comparison to previous indirect data.
At the resonance energy, the branching with proton emission is about $15\%$~\cite{Spillane2007}. Hence, in the stellar environment, in this energy regime, the stellar output is dominated by the $Q$-values and particle generation of carbon fusion reactions with alpha emission. The inclusive reaction rate of a hindrance+resonance model changes drastically as compared to a scenario taking into account only hindrance, but is somewhat comparable to the CF88 model~\cite{Monpribat2022}. However, previously only an average branching into final states was used in stellar simulations. A comprehensive study with 25~\msol{} stars was carried out by Pignatari~\textit{et al.}~\cite{Pignatari2013} applying an arbitrary energy-independent proton-to-alpha scaling of the cross section of $95\%$ to $5\%$, and \textit{vice versa}, with variation of the final abundances of a factor of up to 30 in the subsequent $s$-process. A dominant alpha channel leads to significant decrease of $^{16}$O and $^{23}$Na with increased $^{20}$Ne yields, while a stronger proton channel even changes the main neutron source during the $s$-process to $^{13}$C($\alpha{}$, n).
The work of Jiang~\textit{et al.}~\cite{JiangPRL2013} provides a \sfac{} extrapolation based on partially overlapping resonances~\cite{moldauer1967} due to sparse states in $^{24}$Mg~\cite{davis1981, vanhoy1987} with the aim of overcoming the separation of the cross section into compartments of resonance and background (see \textit{e.g.} Ref.~\cite{aguilera2006PRC}). The density of states drops exponentially with decreasing excitation energy so that single-configuration properties get more pronounced resulting in elevated spaced out fluctuations (likewise seen in Figure~\ref{fig:fracBeck}) and possibly resulting in the significant deviation of the resonance $E_{\mathrm{rel}}=2.14$~MeV or at $E_{\mathrm{rel}}=3.23$~MeV reported in Figure~\ref{fig:fracBeck}. This might be addressed  in nuclear structure studies likewise carried out in first principle calculations in Ref.~\cite{chiba2015PRC}, where resonances in the excitation spectrum of $^{24}$Mg were associated to cluster configurations from isoscalar monopole transitions. We note that the present proton \sfac{}s are largely in agreement with Spillane~\textit{et al.}, albeit with largely improved uncertainty. The detailed comparison with the alpha \sfac{}s, where the uncertainty of gamma detection experiments is \textit{a priori} much higher, depends on the detected exit channels and branching corrections in Ref.~\cite{Spillane2007}.
In summary, we present carbon fusion cross sections from coincident gamma-particle detection in the deep sub-Coulomb barrier energy regime with largely improved uncertainty. Our measurement at $E_{\mathrm{rel}}=2.32$~MeV tends to be below the CF88 extrapolation and in agreement with a previous interpretation of the cross section in the energy regime composed of a hindrance trend with a resonance at $E_{\mathrm{rel}}=2.14$~MeV. Our results at about $E_{\mathrm{rel}}=3.23$~MeV reveal an extremely weak proton branching that might be linked to nuclear structure effects in $^{24}$Mg.
\subsection*{\label{subsec:ack}Acknowledgements}
We thank G.Fremont (GANIL, Caen) for the excellent preparation of the target foils and E. Simpson (Australian National University, Canberra) for the fruitful and productive discussion of our results.
This work of the Interdisciplinary Thematic Institute QMat, as part of the ITI 2021 2028 program of the University of Strasbourg, CNRS and Inserm, was supported by IdEx Unistra (ANR 10 IDEX 0002), and by SFRI STRAT’US project (ANR 20 SFRI 0012) and EUR QMAT ANR-17-EURE-0024 under the framework of the French Investments for the Future Program. MRu, SP, ZsP, PHR and WNC acknowledge support by the UK Science and Technology Facilities Council (STFC) under Grant Nos. ST/P005314/1, ST/V001108/1, ST/L005743/1 and ST/ P005314. PHR acknowledged support from the UK National Measurements System (NMS) Programmes Unit of the UK’s Department for Science, Innovation and Technology. Andromède has benefited from the French research aid  under the program for investment Equipex: ANR-10-EQPX-23. GV acknowledges support from the Consejo Nacional de Ciencia y Tecnologia (CONACyT) reference 2018-000009-01EXTF-00057.
%
%
\bibliographystyle{apsrev4-2}
\bibliography{12C12C_bib}
%
\end{document}